\documentclass[prb,a4paper,twocolumn]{revtex4}
\usepackage{graphicx}

\newcommand{\nc}{\newcommand}

\nc{\be}[1]{\begin{equation}\mbox{$\label{#1}$}}
\nc{\bea}[1]{\begin{eqnarray} \mbox{$\label{#1}$}}
\nc{\Section}[2]{\section{#2}\label{#1}}
\nc{\Bibitem}[1]{\bibitem{#1}}
\nc{\Label}[1]{\label{#1}}

\nc{\eea}{\end{eqnarray}}
\nc{\ee}{\end{equation}}

\nc{\bdm}{\begin{displaymath}}
\nc{\edm}{\end{displaymath}}
\nc{\dpsty}{\displaystyle}
\nc{\bc}{\begin{center}}
\nc{\ec}{\end{center}}
\nc{\ba}{\begin{array}}
\nc{\ea}{\end{array}}
\nc{\bab}{\begin{abstract}}
\nc{\eab}{\end{abstract}}
\nc{\btab}{\begin{tabular}}
\nc{\etab}{\end{tabular}}
\nc{\bit}{\begin{itemize}}
\nc{\eit}{\end{itemize}}
\nc{\ben}{\begin{enumerate}}
\nc{\een}{\end{enumerate}}
\nc{\bfig}{\begin{figure}}
\nc{\efig}{\end{figure}}

\nc{\arreq}{&\!=\!&}
\nc{\arrmi}{&\!-\!&}
\nc{\arrpl}{&\!+\!&}
\nc{\arrap}{&\!\!\!\approx\!\!\!&}
\nc{\non}{\nonumber}
\nc{\align}{\!\!\!\!\!\!\!\!&&}

\def\lsim{\; \raise0.3ex\hbox{$<$\kern-0.75em
      \raise-1.1ex\hbox{$\sim$}}\; }
\def\gsim{\; \raise0.3ex\hbox{$>$\kern-0.75em
      \raise-1.1ex\hbox{$\sim$}}\; }

\nc{\DOT}{\hspace{-0.08in}{\bf .}\hspace{0.1in}}
\nc{\Laada}{\hbox {$\sqcap$ \kern -1em $\sqcup$}}
\nc\loota{{\scriptstyle\sqcap\kern-0.55em\hbox{$\scriptstyle\sqcup$}}}
\nc\Loota{{\sqcap\kern-0.65em\hbox{$\sqcup$}}}
\nc\laada{\Loota}
\nc{\qed}{\hskip 3em \hbox{\BOX} \vskip 2ex}

\nc{\real}{{\rm I \! R}}
\nc{\Z}{{\sf Z \!\!\! Z}}
\nc{\complex}{{\rm C\!\!\! {\sf I}\,\,}}
\def\bigid{\leavevmode\hbox{\small1\kern-3.8pt\normalsize1}}
\def\id{\leavevmode\hbox{\small1\kern-3.3pt\normalsize1}}
\nc{\slask}{\!\!\!/}
\nc{\bis}{{\prime\prime}}
\nc{\pa}{\partial}
\nc{\na}{\nabla}
\nc{\ra}{\rangle}
\nc{\la}{\langle}
\nc{\goto}{\rightarrow}
\nc{\swap}{\leftrightarrow}

\nc{\EE}[1]{ \mbox{$\cdot10^{#1}$} }
\nc{\abs}[1]{\left|#1\right|}
\nc{\at}[2]{\left.#1\right|_{#2}}
\nc{\norm}[1]{\|#1\|}
\nc{\abscut}[2]{\Abs{#1}_{\scriptscriptstyle#2}}
\nc{\vek}[1]{{\rm\bf #1}}
\nc{\integral}[2]{\int\limits_{#1}^{#2}}
\nc{\inv}[1]{\frac{1}{#1}}
\nc{\dd}[2]{{{\partial #1}\over{\partial #2}}}
\nc{\ddd}[2]{{{{\partial}^2 #1}\over{\partial {#2}^2}}}
\nc{\dddd}[3]{{{{\partial}^2 #1}\over
    {\partial #2 \partial #3}}}
\nc{\dder}[2]{{{d #1}\over{d #2}}}
\nc{\ddder}[2]{{{d^2 #1}\over{d {#2}^2}}}
\nc{\dddder}[3]{{d^2 #1}\over
    {d #2 d #3}}
\nc{\dx}[1]{d\,^{#1}x}
\nc{\dy}[1]{d\,^{#1}y}
\nc{\dz}[1]{d\,^{#1}z}
\nc{\dl}[1]{\frac{d\,^{#1}l}{(2\pi)^{#1}}}
\nc{\dk}[1]{\frac{d\,^{#1}k}{(2\pi)^{#1}}}
\nc{\dq}[1]{\frac{d\,^{#1}q}{(2\pi)^{#1}}}

\nc{\bfT}{{\bf T }}

\nc{\cA}{{\cal A}}
\nc{\cB}{{\cal B}}
\nc{\cD}{{\cal D}}
\nc{\cE}{{\cal E}}
\nc{\cG}{{\cal G}}
\nc{\cH}{{\cal H}}
\nc{\cL}{{\cal L}}
\nc{\cO}{{\cal O}}
\nc{\cT}{{\cal T}}
\nc{\cN}{{\cal N}}
\nc{\cR}{{\cal R}}
\nc{\rvac}[1]{|{\cal O}#1\rangle}
\nc{\lvac}[1]{\langle{\cal O}#1|}
\nc{\rvacb}[1]{|{\cal O}_\beta #1\rangle}
\nc{\lvacb}[1]{\langle{\cal O}_\beta #1 |}
\nc{\bb}{\bar{\beta}}
\nc{\bt}{\tilde{\beta}}
\nc{\ctH}{\tilde{\cal H}}
\nc{\chH}{\hat{\cal H}}
\nc{\1}{\aa}
\nc{\2}{\"{a}}
\nc{\3}{\"{o}}
\nc{\4}{\AA}
\nc{\5}{\"{A}}
\nc{\6}{\"{O}}
\nc{\al}{\alpha}
\nc{\g}{\gamma}
\nc{\Del}{\Delta}
\nc{\e}{\textrm{e}}
\nc{\eps}{\epsilon}
\nc{\lam}{\lambda}
\nc{\Om}{\Omega}
\nc{\ve}{\varepsilon}
\nc{\mn}{{\mu\nu}}
\nc{\vp}{\varphi}

\nc{\rf}[1]{(\ref{#1})}
\nc{\nn}{\nonumber \\*}
\nc{\bfB}{\bf{B}}
\nc{\bfv}{\bf{v}}
\nc{\bfx}{\bf{x}}
\nc{\bfy}{\bf{y}}
\nc{\vx}{\vec{x}}
\nc{\vy}{\vec{y}}
\nc{\oB}{\overline{B}}
\nc{\oI}{\overline{I}}
\nc{\oR}{\overline{R}}
\nc{\rar}{\rightarrow}
\nc{\ti}{\times}
\nc{\slsh}{\hskip-5pt/}
\nc{\sm}{Standard~Model~}
\nc{\MP}{M_{\rm Pl}}
\nc{\mpl}{M_{\rm Pl}}
\nc{\tp}{t_{\rm Pl}}

\nc{\pmin}{p_{\rm min}}
\nc{\pmax}{p_{\rm max}}
\nc{\fo}{f_0}
\nc{\foi}{f_{0,i}\,}
\nc{\fop}{f_0^P}
\nc{\fou}{f_0^U}

\nc{\eff}{{\rm eff}}
\nc{\MT}{M_{\rm T}}
\nc{\ML}{M_{\rm L}}
\nc{\kk}{\vek{k}}
\nc{\pp}{{\rm p}}
\nc{\pt}{\partial_t}
\nc{\half}{{1\over 2}}
\nc{\w}{\omega}
\nc{\uhat}{\hat{U}_\w}

\nc{\etal}{\mbox{\it et al.}}
\nc{\ie}{{\it i.e. }}
\nc{\eg}{{\it e.g. }}
\nc{\trh}{T_{\rm RH}}
\nc{\ad}{{a'\over a}}
\nc{\bd}{{b'\over b}}
\nc{\Rd}{{R'\over R}}
\nc{\diag}{{\textrm{diag}}}
\nc{\mato}[1]{\tilde{#1}}
\nc{\sech}{\textrm{sech}}
\nc{\I}{\textrm{I}}
\nc{\II}{\textrm{II}}
\nc{\III}{\textrm{III}}
\nc{\vev}[1]{\langle #1 \rangle}
\nc{\hyp}{\,\; F_{1{\hskip -16pt}2}{\hskip 11pt}}
\nc{\brhom}{\overline{\rho}_M}
\nc{\brho}{\overline{\rho}}
\nc{\rhob}{\overline{\rho}}
\nc{\Pb}{\overline{P}}
\nc{\bH}{\overline{H}}
\nc{\ep}{{1+4\eps}}

\nc{\lcdm}{$\Lambda$CDM}
\nc{\ms}{\langle\sigma\rangle}

\def\smiley{\hbox{\large$\bigcirc$\hspace{-.80em}%
\raise.2ex\hbox{$\cdot\cdot$}\kern-.61em    
\lower.2ex\hbox{\scriptsize$\smile$}}\ }

\def\frowney{\hbox{\large$\bigcirc$\hspace{-.80em}%
\raise.2ex\hbox{$\cdot\cdot$}\kern-.635em
\lower.2ex\hbox{\scriptsize$\frown$}}\ }

\begin{document}

\title{Short-range correlations in  binary alloys: Spin model 
approach to Ag$_c$Pd$_{1-c}$}

\author{I.~Vilja}
\author{K.~Kokko}
\affiliation{Department of Physics and Astronomy, University of Turku, 
FI-20014 Turku, Finland}
\date{\today}

\begin{abstract}
Short-range correlations in Ag-Pd alloys are investigated by analyzing the 
{\em ab initio} total  energy of fcc based random Ag$_c$Pd$_{1-c}$. Since the 
information on the atomic interactions is incorporated in the energetics of
 alloys it is possible with a suitable model, Bethe-Peierls-Weiss model is 
used in the present work, to invert the problem, i.e.\ to obtain information 
on the short-range correlation from the total energy of a random system. As an 
example we demonstrate how site correlations can be extracted from random 
alloy data.  Bethe-Peierls-Weiss model predicts positive first neighbor 
correlator and mixing energy for substitutional face centered cubic (fcc) 
Ag-Pd alloys at low temperature which can be related to the optimal structures 
of Ag$_{0.5}$Pd$_{0.5}$.
\end{abstract}

\maketitle
\section{Introduction}
The low-temperature short-range order of Ag-Pd  has attracted theoretical research for several decades. Results supporting phase separation \cite{Johnson_1990, Takano_1998} as well as ordering \cite{Lu_1991,Muller_2001,Curtarolo_2005,Ruban_2007,Gonis_1983,Takizawa_1989,Abrikosov_1993} have been reported. Due to the low transition temperature predicted for the disorder-order transition there are no direct experimental observation concerning this matter.

In the present work, we reanalyze the Ag-Pd binary alloy system using a spin lattice model beyond the simplest mean field (Weiss) model. We employ Bethe-Peierls-Weiss (BPW) model \cite{BPW}, which, in contrast to Weiss model, incorporates non-vanishing correlators of neighbor sites. Therefore it is suitable for extracting ordering information on the
binary system in hand.

\section{BPW model}
The lattice structure of binary alloys can be modeled by spin (Ising) lattices identifying
spin states $\pm 1$ of a particular site with occupation of a given atom, say A or B. 
For spin model calculations one has to specify the interaction energies between the sites. 
This can be done, for instance, by
fitting the calculated lattice quantities to the corresponding observed or simulated ones. 

In the following we model Ag-Pd by an Ising-lattice with $N$ sites each associated with spin $\sigma_i=\pm 1,\ i=1,\dots N$. We 
identify the spin state $\sigma_i=+1$ with the occupation of the site $i$ by a Ag atom and  
$\sigma_i=-1$ with the occupation of the site $i$ by a Pd atom. The average concentration of
the whole lattice is fixed by requiring that there is $N_+$ sites with $\sigma_i=+1$ and  $N_-$ sites 
with $\sigma_i=-1$ ($N_++N_- = N$). The average concentration of Ag atoms is then $c= N_+/N $ and the
expectation value over all lattice sites is
\be{1}
\langle \sigma\rangle = \frac {\sum_i \sigma_i}{N} = \frac {N_+-N_-}N = 2 c -1.
\ee

Considering only two-particle interactions the state sum is
\be{ZL}
Z=\sum_{\{\sigma_i\}_{i=1}^{N}} e^{-\beta E},
\ee
where the sum is over all spin configurations with constraint  $\sum_i \sigma_i= (2c-1)N$.
The total energy ($E$) is written as the sum of pairwise interaction energies
\be{iE}
E=\frac{1}{2} \sum_{i\ne j} E_{ij} =  \sum_{i > j} E_{ij}.
\ee

An important quantity measuring the ordering of spins is the correlator
\be{corr}
g_{i,j} = \langle \sigma_i\sigma_j\rangle -  \langle \sigma_i\rangle \langle\sigma_j\rangle .
\ee 
Here $g_{i,j}\in [-1,1]$, $g_{i,j}< 0$ ($>0$) corresponds to spins at sites $i$ and $j$ tending to align antiparallel (parallel), whereas the case $g_{i,j}= 0$ corresponds to completely random alignment.

In practise to calculate the state sum in Eq.\ (\ref{ZL}) one has to make further approximation. In the following we use the nearest neighbor (NN) approximation, where only the closest atom sites are 
included in the energy sum (\ref{iE}). 
While Ising models include only two-site interactions, our task requires to include the multi-site interactions
to the model. 
Formally this is done by allowing the Ising model parameters to depend on the concentration $c$, so 
that many-site interactions are effectively and on the average taken into account.

Thus in NN-models only the nearest neighbor sites contribute to the energy and for a fixed concentration
$c$ only effective energy for a site is given. For that purpose we define the pair interaction energy as
\be{NNE}
E_{ij}= \left \{
\begin{array}{ll}
\eps (c) +\Delta\eps (c) \;\sigma_i\sigma_j + \frac 12 \bar\eps(c)\; ( \sigma_i+\sigma_j)& 
\, i,j\ {\rm are\ NN}\\
0 & \, {\rm otherwise}
\end{array}\right .
\ee
Suppose now, that the site $i$ of the Ising-lattice has $\nu_i$ nearest neighbors $j\in \cN_i$. 
This number $\nu_i$ is 
the coordination number of the site $i$. Then the energy of the whole system is
\be{Etot}
E= \frac{1}{2} \sum_{i\ne j} E_{ij}= \frac{1}{2} \sum_{i=1}^N \sum_{j \in \cN_i} E_{ij}
=\frac{1}{2}\sum_{i=1}^N E_i,
\ee
where
\be{Ei}
E_i=\sum_{j \in \cN_i} E_{ij}
\ee
is the effective energy of the site $i$. In this sum there is $\nu_i$ addends and reads for (\ref{NNE}) as
\be{Ei2}
E_i = \nu_i (\eps + \frac 12 \bar\eps\; \sigma_i) 
+ (\Delta\eps\; \sigma_i +\frac 12 \bar\eps) \sum_{j \in \cN_i} \sigma_j.
\ee

To go beyond mean field models 
we treat the system by BPW model \cite{BPW}. 
In BPW model each site $i$ interacts with its $\nu$ nearest 
neighbors \cite{footnote} $i1,\dots i\nu$, which in part interact with
their NN's other that i itself, i.e. with $\nu-1$ sites. The interaction energy of neighbors of $i$ 
are included exactly, whereas the interaction of the neighbors with their neighbors is calculated using
mean field. Thus, instead of Eq.\ (\ref{Ei}) the energy of a site $i$ is
\bea{EnNN}
E_i &=& \sum_{j\in \cN_i} E_{ij} + k \sum_{j\in \cN_i} \sum_{l\in \cN_j,\, l\ne i} \langle E_{jl} 
\rangle_{\sigma_l}\nonumber\\ 
&=& \nu\, \eps + \frac 12 \bar\eps\, \nu\, \sigma_i + (\Delta\eps\, \sigma_i+\frac 12 \bar\eps)
(\sigma_{i1}+\cdots + \sigma_{i\nu})\\ \nonumber
& & +k\, (\nu-1) [\nu\, \eps + (\Delta\eps \ms+\frac 12 \bar\eps) (\sigma_{i1}+\cdots + \sigma_{i\nu}) \\ \nonumber 
& & + \frac 12 \nu \bar\eps \ms],
\eea
where $k$ is a combinatorial parameter used to obtain specific thermodynamic quantities by differentiation and the choice $k=1$ corresponds to the normal BPW model. For Ag-Pd the pair interactions just at the first two coordination shells are needed to qualitatively understand the ordering energy \cite{Ruban_2007}.

For $\nu$ NNs of site $i$ we denote the number of $\sigma_{ij}=+1$ by $n_+$ and 
the number of $\sigma_{ij}=-1$ by $n_-$, so that $n_+ +n_- = \nu$ and 
\bea{EBPW}
E_i &=& E(\sigma_i, n_+ )\equiv 
\nu\, \eps + \frac 12 \bar\eps\, \nu\, \sigma_i + (\Delta\eps\, \sigma_i+\frac 12 \bar\eps)
(n_+ - n_-)\nonumber \\
& & +\ k\, (\nu-1) [\nu\, \eps + (\Delta\eps \ms+\frac 12 \bar\eps) (n_+-n_-) + 
\frac 12 \nu \bar\eps \ms]\nonumber\\
&=& \nu \left [\eps -\frac 12 \bar\eps + k (\nu-1)\left \{ \eps- \Delta\eps \ms - \frac 12\,\bar\eps\, 
(1-\ms )\right \}
\right ] \nonumber \\
& & + \nu [\frac 12 \bar\eps -\Delta\eps ] \sigma_i\nonumber \\
& & +\left [\Delta\eps\, \sigma_i + \frac 12 \bar\eps + k (\nu-1) \left \{\Delta\eps\ms + 
\frac 12\bar\eps\right\}
\right ] 2\, n_+.
\eea

When $n_+\ll N_+$ and $\nu\ll N$ we may approximate that configuration of the $\nu$ neighbors of a fixed
site $i$ is effectively independent on all other sites. This is the main idea of Bethe-Peierls approximation \cite{Robertson}. However, BPW-approximation
is widely used to model short-range correlation in various statistical systems \cite{VJ, VJS}. The validity of this approximation is however not clear, as there is no known error estimation method; PBW model has an uncontrolled error. Anyway, with this assumption of effective independence of neighbors of separate sites leads to partition function
\be{ZBPW}
Z= \sum_{ \{\sigma_k\}_{k=1}^N } \prod_{i=1}^N e^{-\frac 12 \beta E_i}=
\left (\begin{array}{c} N\\N_+ \end{array}\right ) Z_1 (+1)^{N_+} Z_1(-1)^{N_-},
\ee
where $\beta=(k_{\rm B}T)^{-1}$, $k_{\rm B}$ is the Boltzmann constant and $T$ temperature. 

The effective one site (neighbor) partition function for spin $\sigma$
can be calculated using grand canonical ensemble of its
neighbors. We write
\be{Z1}
Z_1(\sigma) = \sum_{n_+=0}^{\nu} \left (\begin{array}{c} \nu\\n_+ \end{array}\right )
e^{\mu' n_+} e^{-\frac 12 \beta E(\sigma, n_+)}.
\ee
The chemical potential of a site $\mu'$ has to be related to the overall concentration $c$,
as $N_+$ is still fixed by overall concentration condition $c=N_+/N$.

Here the partition function $Z$ is a grand canonical partition function with respect
to the number of $\sigma = +1$ sites of the system but canonical one with respect
to the total number of sites. That is, the partition function is related to
grand potential
$$
\Omega = -\frac 1\beta \ln Z.
$$
Thus when we later turn to use Helmholtz free energy (i.e fixed concentration) we have to make
the appropriate Legendre transformation. 

The thermodynamical quantities are to be calculated from the logarithm of the partition function, i.e.
\bea{logZBPW}
\ln Z &=& \ln \left ( \begin{array}{c} N\\N_+ \end{array}\right )\nonumber \\ 
&-&  \frac 12\, N\, \beta [E_0 + 
c \Delta E(+1) + (1-c)\Delta E(-1)]\nonumber \\
&+& \ N\left \{ \nu\, c\, \ln\left [1+z\,e^{-\beta \tilde E_+}\right.\right ]\nonumber \\ 
&+& \left. \nu\, (1-c)\,
\ln \left [1+z\, e^{-\beta \tilde E_-}\right ]\right \} 
\eea 
with
\bea{Es}
E_0&=& \nu \left [\eps - \frac 12 \bar\eps\right.\nonumber \\ 
&+&\left. k (\nu-1)\left \{ \eps- \Delta\eps \ms - 
\frac 12\,\bar\eps\, (1-\ms )\right \}
\right ]\nonumber\\
&=& \nu \left [ \nu_*\eps -\nu_*\frac 12 \bar\eps + k (\nu-1)\left \{ \frac 12 \bar\eps
-\Delta\eps \right\} \ms\right ]\nonumber\\
\Delta E(\sigma) &=& \Delta E\; \sigma = \nu [\frac 12 \bar\eps -\Delta\eps ] \sigma , \nonumber\\
\tilde E(\sigma) &=& \Delta\eps\, \sigma +\frac 12 \bar\eps + k (\nu-1) \left \{\Delta\eps\ms + 
\frac 12 \bar\eps\right\} \nonumber\\
&=& \Delta\eps \sigma + \nu_*\frac 12 \bar\eps + k (\nu-1)\Delta\eps \ms \nonumber
\eea
where $z=e^{\mu '}$ is the one-site fugacity
\bea{z}
z^{-1}&=& \frac { \sqrt {\ms^2 + (1-\ms^2)e^{-\beta (\tilde E_+-\tilde E_-)}}
-\ms}{1+\ms}\; e^{-\beta \tilde E_-}
\nonumber\\
&=& 
\frac { \sqrt {\ms^2 + (1-\ms^2)e^{-2 \beta\Delta\eps}} -\ms}{1+\ms}\;
e^{- \beta \tilde E_- },
\eea
$\tilde E_\pm = \tilde E (\pm 1)$, and $\nu_*=1+k(\nu -1)$.
The internal energy reads
\bea{UBPW}
U&=& -\frac{\partial \ln Z}{\partial\beta}=\frac 12\, N\, [E_0 + (2c-1)\Delta E] \nonumber\\
&& +\ N\left [ \frac {c\,\nu\,\tilde E_{+} }{1+z^{-1} e^{\beta \tilde E_{+}}}+
\frac {(1-c)\,\nu\,\tilde E_{-}}{1+z^{-1} e^{\beta \tilde E_{-}}}\right ].
\eea

Now we turn to use fixed concentration, whence the Helmholtz free energy is given by
\be{OF}
F=\Omega + \frac {\mu'}\beta\frac {\partial\ln Z}{\partial\mu'}= -\frac1\beta \ln Z + \frac 1\beta N_+\nu
\ln z
\ee
which can be expressed as energy density.
The formula for the entropy can be given after that in a standard way:
\bea{SBPW}
T\,S &=& U-F = U+ \frac 1\beta \ln Z- \frac 1\beta  N_+\nu \ln z \nonumber\\
&=& T\,S_0 + \frac {N \nu}\beta [c I_+ + (1-c)I_--  c \ln z], 
\eea
where
\be{I}
I_\pm  = \frac {\beta \tilde E_\pm }{1+z^{-1} e^{\beta \tilde E_\pm}}+
\ln [1+z\,e^{-\beta \tilde E_\pm}].
\ee
Further, the entropy per site is
\be{sBPW0}
s= s_0 + k_B\,  \nu\,[c I_+ + (1-c)I_-] - k_B\, \nu\, c \ln z
\ee
and the Helmholtz free energy per site reads as
$$
f=-\frac1{N \beta}\, \ln Z + \frac 1\beta \nu\, c \ln z.
$$
The mixing energy per atom can be constructed straightforwardly as
$$
u_{mix}(c)= u(c) - c\, u(1)-(1-c)\,u(0).
$$

By differentiation of (\ref{logZBPW}) with respect to energy parameters $\bar\eps$ and $\Delta\eps$ we find
\bea{apu3}
g_{i,\, i1} &=& [1- \ms^2]\left [
\frac 1{1+z^{-1}e^{\beta \tilde E_{+}}}-\frac 1{1+z^{-1}e^{\beta \tilde E_{-}}}\right ].
\eea
Thus BPW model clearly allows correlation between a site and its neighbors.
\section{Results}
Next step is to determine the BPW parameters $\eps,\ \bar\eps$ and $\Delta\eps$ by
fitting the internal energy to the calculated total energy of Ag$_c$Pd$_{1-c}$ alloys. The reference data we are using are the total energies from Ref.~\onlinecite{RKVK} calculated for Ag$_c$Pd$_{1-c}$ using the exact muffin-tin orbitals method \cite{vitos2007,vitos2001} within the coherent potential approximation \cite{soven1967,gyorffy1972}, i.e.\ corresponding to the mean field approximation (Table \ref{t1}). Because the {\it ab initio} energies in Table \ref{t1} refer to the {\em non-correlated} Ag-Pd system at 0~K temperature
we have to use in the fitting procedure the non-correlated BPW energy obtained by differentiation with respect to $k$ or equivalently from Eq.\ (\ref{EnNN}) as 
$$
u= \frac 1{2N} \sum_{i=1}^N \langle E_i \rangle,
$$
assuming that $\langle \sigma_i\sigma_j\rangle =
\langle \sigma_i\rangle\langle \sigma_j\rangle = \ms^2$ ($i\ne j$).
We find the non-correlated internal energy $u_{nc}$ to be up to normalization same as in the Weiss model,
\be{uncs}
u_{nc}(\ms ) = \frac 12 \nu\nu_* [ \eps + \bar\eps \ms + \Delta\eps \ms^2 ].
\ee
This can be expressed as a 
function of concentration as
\be{uncc}
u_{nc}(c ) = \frac 12 \nu\nu_* [ \eps_c + \bar\eps_c c + \Delta\eps_c c^2],
\ee
where the parameters are related by
\bea{rel}
\Delta\eps &=& \frac 14 \Delta\eps_c,\nonumber\\
\bar\eps &=& \frac 12 (\Delta\eps_c+\bar\eps_c),\\
\eps &=& \eps_c+\frac 12 \bar\eps_c + \frac 14 \Delta\eps_c.\nonumber
\eea

\begin{table}[ht]
\begin{center}
\begin{tabular}{lc}
\hline
\hline
$c$& $u_{nc}(c)$[Ry]\\
\hline
0.00&  -10084.720387\\
0.05& -10111.731714\\
0.10& -10138.743066\\
0.20& -10192.765913\\
0.40& -10300.812124\\
0.50& -10354.835440\\
0.60& -10408.858240\\
0.70& -10462.880386\\
0.80& -10516.901915\\
0.90& -10570.922942\\
1.00& -10624.943741\\
\hline
\hline
\end{tabular}
\end{center}
\caption{The {\it ab initio} internal energy of Ag$_c$Pd$_{1-c}$ at 0 K as obtained in Ref.~\onlinecite{RKVK}.}\label{t1}
\end{table}

To fit the Eq.\ (\ref{uncc}) to the numerical data shown in Table~\ref{t1} is not an unambiguous procedure. Several justified strategies can be introduced to define the concentration dependence of the parameters $\eps_c$, $\bar\eps_c$, and $\Delta\eps_c$. We study four different fitting schemes and compare their results with the data obtained from impurity calculations for Ag-Pd system. In three first fitting schemes we take $\eps_c$ and $\bar\eps_c$ to be constants whereas different boundary conditions for $\Delta\eps_c$ are introduced.
\begin{itemize}
\item[i)] 
The choice $\Delta\eps_c(1)=\Delta\eps_c(0)=0$ leads to the correct linear form of the total energy in the limit of noninteracting atoms. This leads to tenth degree interpolation polynomial for $\Delta\eps_c(c)$.
\item[ii)] 
Since the term $c^2 \Delta\eps_c(c)$ vanish at $c=0$ the above constraint can be lifted to $\Delta\eps_c(1)=0$. This leads to ninth degree interpolation polynomial for $\Delta\eps_c(c)$.
\item[iii)] 
Fitting of eight degree polynomial to the data in Table~\ref{t1} and interpreting the two terms of lowest degree as $\eps_c$ and $\bar\eps_c c$, now  neither $\Delta\eps_c(0)= 0$ nor $\Delta\eps_c(1)= 0$. 
\item[iv)] 
Redlich-Kister (RK) parameterization \cite{RK} for Eq.\ (\ref{uncc})
\be{unRK}
u_{nc}(c ) = \frac 12 \nu\nu_* [ \eps_{RK} + \bar\eps_{RK} c + c (1-c)\Delta\eps_{RK}(c)]. \nonumber
\ee
Part of the linear term in (\ref{uncc}) is included to the RK-interaction term. Note, that here
$\eps_{RK}$ and $\bar\eps_{RK}$ are constants. Thus $\eps_{c}=\eps_{RK}$, $\bar\eps_c = \bar\eps_{RK}+
\Delta\eps_{RK}$ and 
$\Delta\eps_c(c)= -\Delta\eps_{RK}$. The RK-parameterization is symmetric in the exchange of Pd$\leftrightarrow$Ag.
\end{itemize}
To decide which one of the above fitting procedures describes best the energy of Ag-Pd alloy we compare the parameters $\eps(c)$, $\bar\eps(c)$, and $\Delta\eps(c)$ at both ends ($c=0$, $c=1$) with the corresponding calculated data of Ag-Pd. The reference parameters are shown in Table~\ref{t5}. 
\begin{table}[ht]
\begin{center}
\begin{tabular}{lcc}
\hline
\hline
 & $c=0$ & $c=1$\\
\hline
$\eps_{\rm comp}$ [Ry]& -143.8171426 & -143.8171231\\
$\bar\eps_{\rm comp}$ [Ry]& -3.7515816 & -3.7516004\\
$\Delta\eps_{\rm comp}$ [$\mu$Ry]& 0.0& 60.417\\
\hline
\hline
\end{tabular}
\end{center}
\caption{Energy parameters $\eps,\ \bar\eps$ and $\Delta\eps$ obtained from Refs. \onlinecite{HSZD} and \onlinecite{Skriver_db}.}\label{t5}
\end{table}
\begin{table}[ht]
\begin{center}
\begin{tabular}{lccc}
\hline
\hline
$c$ & $\eps$[Ry]  & $\bar\eps$[Ry]& $\Delta\eps$[$\mu$Ry]\\
\hline
0.00& -143.817139& -3.75158336& -5.555558\\
0.05& -143.817143& -3.75159123& -9.490742\\
0.10& -143.817147& -3.75159910& -13.425927\\
0.20& -143.817154& -3.75161484& -21.296297\\
0.40& -143.817159& -3.75162327& -25.513510\\
0.50& -143.817159& -3.75162367& -25.712963\\ 
0.60& -143.817153& -3.75161104& -19.396219\\
0.70& -143.817144& -3.75159374& -10.747355\\
0.80& -143.817135& -3.75157664& -2.199074\\
0.90& -143.817128& -3.75156189& 5.178326\\
1.00& -143.817123& -3.75155107& 10.587963\\
\hline
\hline
\end{tabular}
\end{center}
\caption{Energy parameters $\eps$, $\bar\eps$, and $\Delta\eps$ as a function 
of concentration according to the fitting scheme ($iii$). }\label{t3}
\end{table}

\begin{figure}[th]
\leavevmode
\centering 
\vspace*{68mm}
\begin{picture}(0,60)(0,500)
\put(-117, 460.0){$f_{mix}$[Ry]}
\put(110,427){$c$}
\includegraphics{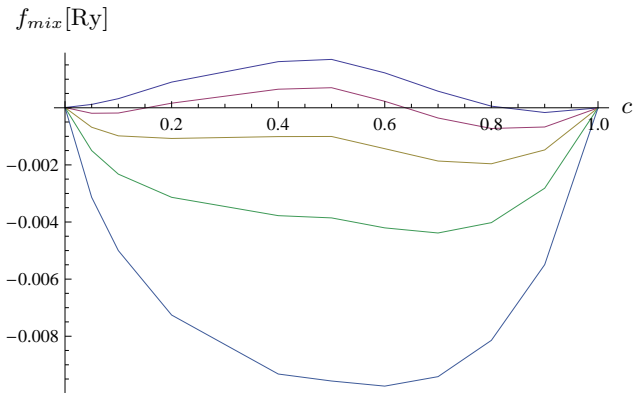}
\end{picture}
\caption{(Color online) Helmholtz free mixing energy per site $f_{mix}$ of Ag$_c$Pd$_{1-c}$ at 
$T=1{\rm K},\ 20{\rm K},\ 50{\rm K},\ 100{\rm K},\ 200{\rm K}$ (from up to down) 
.}
\label{f3}
\end{figure}

\begin{figure}[th]
\leavevmode
\centering 
\vspace*{68mm}
\begin{picture}(0,60)(0,500)
\put(-117, 470.0){$c_{i,i1}$}
\put(110,329){$c$}
\includegraphics{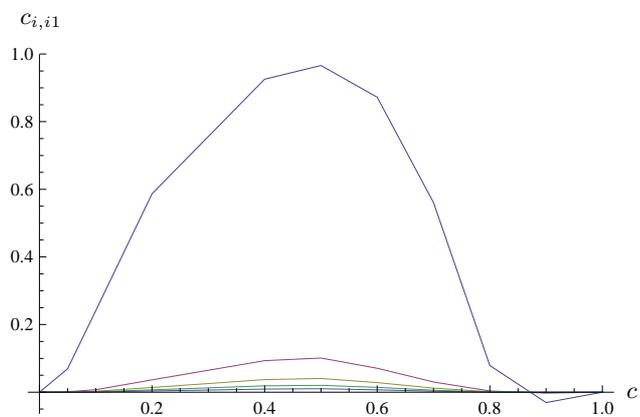}
\end{picture}
\caption{(Color online) Neighbor correlators of Ag$_c$Pd$_{1-c}$ at various temperatures
$T=1{\rm K},\ 20{\rm K},\ 50{\rm K},\ 100{\rm K},\ 200{\rm K}$ (from up to down).}
\label{f7}
\end{figure}

It turns out that the fitting procedure ($iii$) gives the best overall agreement with the {\em ab initio} impurity data. The BPW parameters obtained from the fit ($iii$) are shown in Table \ref{t3}. Using the BPW parameters various thermodynamic and statistical quantities can be calculated for Ag$_c$Pd$_{1-c}$. Figures \ref{f3} and \ref{f7} show the mixing energy and the neighbor correlator at different temperatures. 

At low temperatures the mixing energy and the neighbor correlator are positive suggesting that the BPW multisite interactions drive substitutional fcc Ag-Pd to favor an atom and its neighbors to be of the same type. This suggests phase separation or segregation tendency for Ag-Pd alloys within substitutional fcc structures at low temperatures in agreement with the results of Refs.~\onlinecite{Johnson_1990} and \onlinecite{Takano_1998}. However, as the recent investigations of Delczeg-Cirjak et al. \cite{Delczeg-Czirjak2011} show the ground state structure of Ag$_{0.5}$Pd$_{0.5}$ is not a substitutional fcc type structure but the L1$_1$ type structure with $c/a$ larger than its ideal value. Here $c$ and $a$ are the lattice parameters in hexagonal representation. In conventional cubic coordinate system $c$ axis is along the [111] direction and the lattice parameter $a$ is in the (111) plane. For ideal fcc structure $c/a=\sqrt{6}$, where $c$ is the diagonal of the conventional cube and $a$ is the nearest neighbor distance. Since the L1$_1$ structure is composed of alternating Ag and Pd (111) layers the increasing of $c/a$ compared to the ideal value means, that the nearest neighbor distance of unlike pairs becomes larger than that of like pairs. This shifts more weight on the like-pair interaction than on the mixed-pair interaction. Therefore, the results of BPW model are consistent with the predicted tendency of Ag$_{0.5}$Pd$_{0.5}$ at lower temperatures to order in L1$_1$ structure with increased $c/a$.

At medium temperatures (50 -- 100~K) the mixing energy is negative but the maximum at about $c=0.5$ suggests a slight tendency of phase separation. At high temperatures the BPW mixing energy approaches the experimental mixing enthalpy \cite{Hultgren1973} which has its minimum at about $c=0.6$.

\section{Summary}
We have used the Bethe-Peierls-Weiss model to investigate the effect of multisite interactions on the ordering in Ag-Pd alloys. The mean field {\em ab initio} data has been used to determine the parameters of the BPW model. The BPW mixing energy and neighbor correlator for a substitutional fcc structure of Ag-Pd are positive at low temperatures supporting the stability of the L1$_1$ structure of Ag$_{0.5}$Pd$_{0.5}$ with elongation along the hexagonal [111] axis.

\begin{thebibliography}{X}

\bibitem{Johnson_1990} R. A. Johnson, Phys. Rev. B {\bf 41}, 9717 (1990). 
\bibitem{Takano_1998} N. Takano, A. Yoshikawa, and F. Terasaki, Solid  State Commun. {\bf 107}, 213 (1998).
\bibitem{Lu_1991} Z. W. Lu, S.-H. Wei, A. Zunger, S. Frota-Pessoa, and L. G. Ferreira, Phys. Rev. B {\bf 44}, 512 (1991).
\bibitem{Muller_2001} S. M{\"u}ller and A. Zunger, Phys. Rev. Lett. {\bf 87}, 165502 (2001).
\bibitem{Curtarolo_2005} S. Curtarolo, D. Morgan, and G. Ceder, Comput Coupling Phase Diagrams Thermochem. {\bf 29}, 163 (2005).
\bibitem{Ruban_2007} A. V. Ruban, S. I. Simak, P. A. Korzhavyi, and B. Johansson, Phys. Rev. B {\bf 75}, 054113 (2007).
\bibitem{Gonis_1983} A. Gonis, W. H. Butler, and G. M. Stocks, Phys. Rev. Lett. {\bf 50}, 1482 (1983).
\bibitem{Takizawa_1989} S. Takizawa, K. Terakura, and T. Mohri, Phys. Rev. B {\bf 39} 5792 (1989).
\bibitem{Abrikosov_1993} A. I. Abrikosov and H. L. Skriver, Phys. Rev. B {\bf 47}, 16532 (1993).
\bibitem{BPW} P. R. Weiss, Phys. Rev. {\bf 74}, 1493 (1948).
\bibitem{footnote} We suppose here again, that each site has same coordination number.
\bibitem{Robertson} H. S. Robertson, {\em Statistical Thermophysics} (Prentice-Hall, New Jersey, 1993).
\bibitem{VJ} T. Vojta and W. John, J. Phys.: Condens. Matter {\bf 5}, 57 (1993).
\bibitem{VJS} T. Vojta, W. John and M. Screiber, J. Phys.: Condens. Matter {\bf 5}, 4989 (1993). 
\bibitem{RKVK} M. Ropo, K. Kokko, L. Vitos and J. Koll\' ar, Phys. Rev. B {\bf 71}, 45411 (2005).
\bibitem{vitos2007} L. Vitos, {\em Computational Quantum Mechanics for Materials Engineers: The EMTO Method and Applications}, Engineering Materials and Processes Series (Springer-Verlag, London, 2007).
\bibitem{vitos2001} L. Vitos, I. A. Abrikosov, and B. Johansson, Phys. Rev. Lett. {\bf 87}, 156401 (2001).
\bibitem{soven1967} P. Soven, Phys. Rev. {\bf 156}, 809 (1967).
\bibitem{gyorffy1972} B. L. Gy\"orffy, Phys. Rev. B {\bf 5}, 2382 (1972).
\bibitem{RK} O. Redlich and A. T. Kister, Ind. Eng. Chem. {\bf 40}, 345 (1948).
\bibitem{HSZD} T. Hoshino, W. Schweika, R. Zeller, and P. H. Dederichs, Phys. Rev. B {\bf 47}, 5106 (1993).
\bibitem{Skriver_db} www.imprs-am.mpg.de/summerschool2003/skriver.pdf
\bibitem{Delczeg-Czirjak2011} E. Delczeg-Czirjak, et al., Unpublished.
\bibitem{Hultgren1973} R. Hultgren et al., {\em Selected Values of the Thermodynamic Properties of Binary Alloys} (American Society for Metals, Metals Park, Ohio, 1973).



\end{thebibliography}
\end{document}